\documentstyle[here,epsfig]{mn2e}
\def\simlt{\mathrel{\rlap{\lower 3pt\hbox{$\sim$}}\raise 2.0pt\hbox{$<$}}}
\def\simgt{\mathrel{\rlap{\lower 3pt\hbox{$\sim$}} \raise 2.0pt\hbox{$>$}}}

\def\gtsima{$\; \buildrel > \over \sim \;$}
\def\ltsima{$\; \buildrel < \over \sim \;$}
\def\gtrsim{\lower.5ex\hbox{\gtsima}}
\def\lesssim{\lower.5ex\hbox{\ltsima}}
\def\url#1{{\ttfamily\def\/{/\diskretionary{}{}{}}#1}}

\newcommand{\q}{\begin{equation}}
\newcommand{\qa}{\begin{eqnarray}}
\newcommand{\qs}{\begin{eqnarray*}}
\newcommand{\nq}{\end{equation}}
\newcommand{\nqa}{\end{eqnarray}}
\newcommand{\nqs}{\end{eqnarray*}}

\begin{document}

\title[Formation of the massive stars around SgrA$^\ast{}$]{{\it In situ} formation of the massive stars around SgrA$^\ast{}$}
\author[Mapelli et al.]
{M. Mapelli$^{1}$, T. Hayfield$^{1,2}$, L. Mayer$^{1,2}$, J. Wadsley$^{3}$
\\
$^{1}$ Institute for Theoretical Physics, University of Z\"urich, Winterthurerstrasse 190, CH-8057, Z\"urich, Switzerland; {\tt mapelli@physik.unizh.ch}\\
$^{2}$ Institute of Astronomy, ETH Z\"urich, ETH Honggerberg HPF D6, CH-8093, Z\"urich, Switzerland\\
$^{3}$Department of Physics and Astronomy, McMaster University, Hamilton, ON L8S 4M1, Canada\\
}

\maketitle \vspace {7cm }

\begin{abstract}
The formation of the massive young stars surrounding  SgrA$^\ast{}$ is still an open question. In this paper, we simulate the infall of an isothermal, turbulent 
molecular cloud towards the Galactic Centre (GC). 
As it spirals towards the GC, the molecular cloud forms a small and dense disc around SgrA$^\ast{}$.
Efficient star formation (SF) is expected to take place in such a dense disc. We model SF by means of sink particles.
At $\sim{}6\times{}10^5$ yr, $\sim{}6000\,{}M_\odot{}$ of stars have formed, and are confined within a thin disc with inner and outer radius of 0.06 and 0.5 pc, respectively. 
Thus, this preliminary study shows that the infall of a molecular cloud is a viable scenario for the formation of massive stars around SgrA$^\ast{}$. Further studies with
more realistic radiation physics and SF will be required to better constrain this intriguing scenario.
\end{abstract}

\section{Introduction}
The origin of young massive stars which crowd the Galactic Centre (GC) has been a puzzle for a long time. Most of the massive stars observed in the central parsec reside in one or
perhaps two rotating discs (Genzel et al. 2003, hereafter G03; Paumard et al. 2006, hereafter P06). 
These discs have well-defined inner ($r_{in}\sim{}0.04$) and outer radii ($r_{out}\sim{}0.5$ pc). In fact, no OB stars have been found at a distance larger than $\sim{}0.5$ pc (P06) from SgrA$^\ast{}$, the source identified with the super massive black hole (SMBH). Similarly, the $S$ stars observed at distances $\lesssim{}0.02$ pc have randomly oriented motions and do not belong to the discs (G03; Ghez et al. 2005; Eisenhauer et al. 2005).
The massive stars inside the discs are young ($6\pm{}2$ Myr, P06) and must have formed over a short period ($<2$ Myr, P06). Their estimated initial mass function (IMF) is heavier than Salpeter's one (P06). The total mass in the discs cannot exceed $1.5\times{}10^4\,{}M_\odot{}$, but is more likely of the order of $5\times{}10^3\,{}M_\odot{}$ (P06). 

Such stars cannot have formed {\it in situ}
in 'normal' conditions, as the tidal forces exerted from the SMBH would have disrupted
 the parent molecular cloud (Levin \& Beloborodov 2003; G03). Thus, an alternative scenario has been proposed, according to which a young cluster spiraled towards the GC and deposited its stars around SgrA$^\ast$ (Gerhard 2001; McMillan \& Portegies Zwart 2003; 
Kim \& Morris 2003; 
G\"urkan \& Rasio 2005; Fujii et al. 2008).  However, even the latter scenario suffers from various shortcomings, such as the premature disruption of the cluster and the excessively 
long dynamical friction time. These problems have only been partially solved by assuming that the original clusters 
host intermediate-mass black holes (Portegies Zwart et al. 2006), or by using new computational schemes (Fujii et al. 2008).
\begin{figure}
\center{{
\epsfig{figure=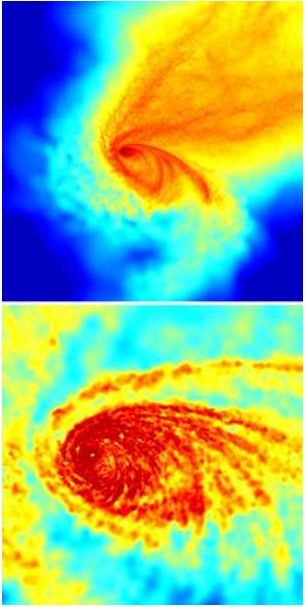,width=4.0cm}
}}
\caption{\label{fig:fig1} Upper panel: density map of gas, projected along the $x-$axis at $t=3.3\times{}10^5$ yr. The frame measures 20 pc per edge and the centre is at 2 pc from the SMBH. 
The density goes from $2.23\times{}10^{-2}$ (blue) to $4.45\times{}10^3\,{}M_\odot{}$ pc$^{-2}$ (red) in logarithmic scale.
Lower panel: zoom of the upper panel.  The frame measures 1.5 pc per edge and the centre is at 0.3 pc from the SMBH. The density goes from $7.05\times{}10^{1}$ to $5.60\times{}10^3\,{}M_\odot{}$ pc$^{-2}$ in logarithmic scale.
}
\end{figure}
On the other hand, the problem of  tidal forces exerted by the SMBH can be overcome if,
 at some point 
in the past, 
a dense gaseous disc existed around SgrA$^\ast{}$. Such a disc could have formed due to the infall and tidal disruption of a molecular cloud. If the density in the disc was high enough, 
it might have become unstable to fragmentation and formed stars (Levin \& Beloborodov 2003; G03; Goodman 2003; Milosavljevic \& Loeb 2004; Nayakshin \& Cuadra 2005; Alexander et al. 2008; Collin \& Zahn 2008). The absence of massive stars at distances $>0.5$ pc from SgrA$^\ast{}$ supports this idea of {\it in situ} formation (Nayakshin \& Sunyaev 2005; P06). This scenario is also favoured by the existence of two giant molecular clouds within $\sim{}20$ pc from the dynamical centre of our Galaxy (Solomon et al. 1972). One of these two clouds (named M$-0.13-0.08$) is also highly elongated toward SgrA$^\ast{}$ and has a 'finger-like' extension pointing in the direction of the circumnuclear disc (Okumura et al. 1991; Ho et al. 1991; Novak et al. 2000, and references therein).
Nayakshin, Cuadra \& Springel (2007, hereafter NCS07) simulated star formation (SF) in a gaseous disc around SgrA$^\ast{}$, and found encouraging results for this scenario. However, NCS07 assume that the gaseous disc was already in place when it started forming stars, and do not consider the process which lead to the formation of the disc itself. In this paper, we simulate the infall of a molecular cloud toward SgrA$^\ast{}$ and we study the formation of a dense gaseous disc around the SMBH. We use a simple phenomenological recipe to simulate the SF process in the disc. Our results indicate that the infall of a molecular cloud is a viable scenario for the formation of massive stars around SgrA$^\ast{}$.

\section{Models and simulations}
We ran N-body/Smoothed Particle Hydrodynamics (SPH) simulations of a molecular cloud evolving in a potential dominated by the SMBH.
The SMBH is represented by a sink particle, with initial mass $M_{\rm BH}=3.5\times{}10^6\,{}M_\odot{}$ (Ghez et al. 2003), sink radius $r_{acc}=5\times{}10^{-3}$ pc and softening radius $\epsilon{}=1\times{}10^{-3}$ pc. We also add a rigid potential, according to a density distribution $\rho{}(r)=\rho{}_0\,{}(r/c)^{-\alpha{}}$, where $\rho{}_0=1.2\times{}10^6\,{}M_\odot\textrm{ pc}^{-3}$, $c=0.39$ pc, and $\alpha{}=1.4$ at $r<c$ and $=2.0$ at $r>c$ (G03).

The cloud used in this experiment is spherical with a radius of 15 pc, a
mass of $4.3\times{}10^4\,{}M_\odot{}$, and a temperature\footnote{A temperature of 10-30 K is consistent with the one predicted for dense gas in regions of moderate star formation (Spaans \& Silk 2000). Test simulations show that our results do not change significantly for isothermal clouds with temperature up to $\sim{}50$~K.} of 10~K. It is seeded with supersonic turbulent velocities and
marginally self-bound. 
A total of  $2\,{}155\,{}660$ particles were
employed, thus the particle mass is 0.02 $M_\odot{}$.
 To simulate interstellar
turbulence, the velocity field of the cloud was generated on a grid as a divergence-free
Gaussian random field with an imposed power spectrum P($k$), varying as $k^{-4}$.
This
yields a velocity dispersion $\sigma{}(l)$, varying as $l^{1/2}$, chosen to agree with
the Larson scaling relations (Larson 1981). The velocities were then interpolated from the
grid to the particles. Finally, the condition that the cloud be marginally
self-bound gives a normalization for the global velocity dispersion of $3.8\,{}{\rm km}\,{}{\rm s}^{-1}$.
\begin{figure*}
\center{{
\epsfig{figure=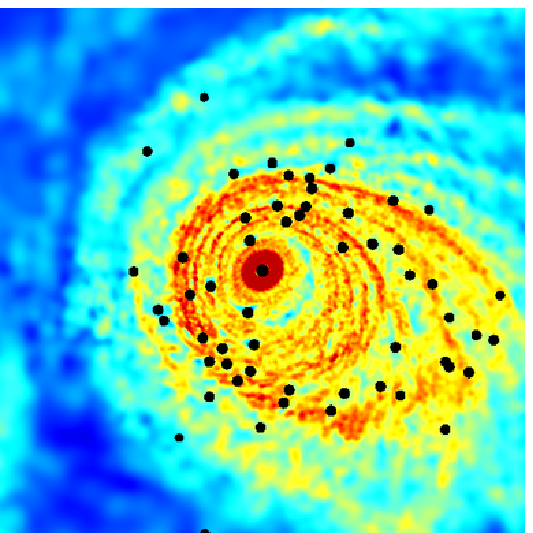,height=4.5cm}
\epsfig{figure=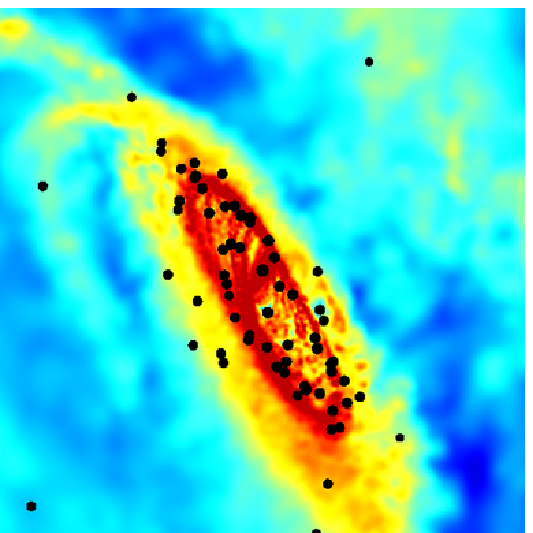,height=4.5cm}
\epsfig{figure=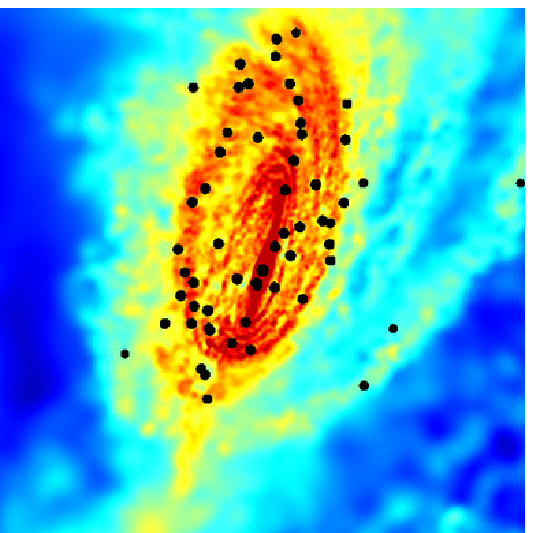,height=4.5cm}
}}
\caption{\label{fig:fig2} Density map of gas in the central disc, projected along the $x-$ (left-hand panel), $y-$ (central panel) and $z-$axis (right-hand panel) at $t=4\times{}10^5$ yr. The frames measure 1 pc per edge. The density goes from $2.23\times{}10^{2}$ to $2.81\times{}10^4\,{}M_\odot{}$ pc$^{-2}$ in logarithmic scale. The superimposed filled black circles are the sink particles. The sink particle at the centre of the frames is the SMBH. 
}
\end{figure*}
\begin{figure*}
\center{{
\epsfig{figure=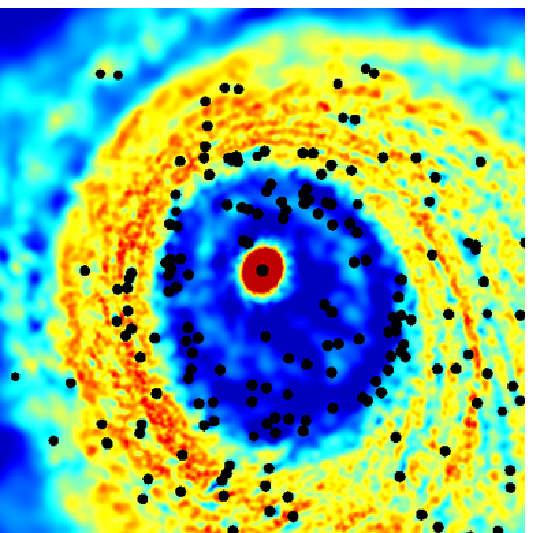,height=4.5cm}
\epsfig{figure=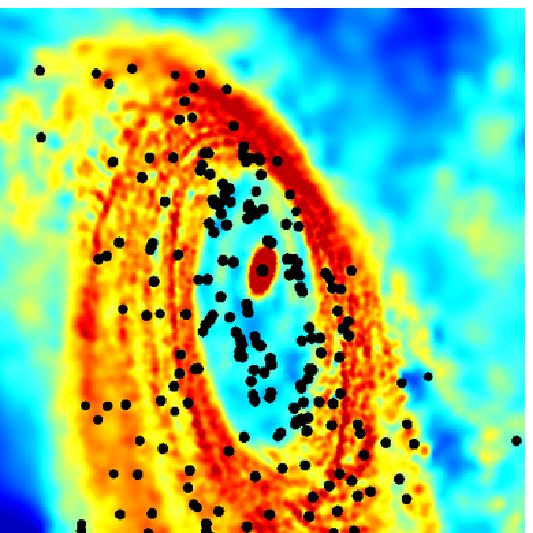,height=4.5cm}
\epsfig{figure=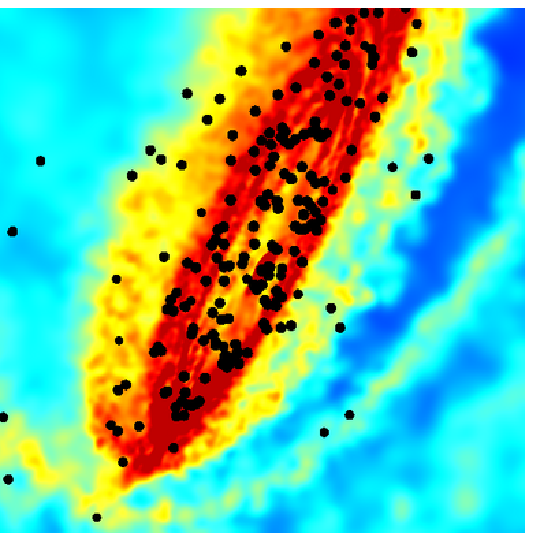,height=4.5cm}
}}
\caption{\label{fig:fig3} Density map of gas in the central disc, projected along the $x-$ (left-hand panel), $y-$ (central panel) and $z-$axis (right-hand panel) at $t=5\times{}10^5$ yr. The frames measure 1 pc per edge. The density goes from $2.23\times{}10^{2}$ to $1.12\times{}10^4\,{}M_\odot{}$ pc$^{-2}$ in logarithmic scale. The superimposed filled black circles are the sink particles. The sink particle at the centre of the frames is the SMBH. 
}
\end{figure*}
The centre-of-mass of the cloud is initially at 25 pc from the SMBH. The orbit of the cloud was chosen so that the impact parameter with respect to the SMBH is $10^{-2}$ pc and the initial
 velocity is one tenth of the escape velocity from the SMBH at the initial distance (i.e. the orbit is bound and highly eccentric). As the tidal density  at $\sim{}25$ pc from the GC is higher than the average initial density of our cloud, our marginally bound cloud must have formed further out and then migrated closer to the GC. Various processes could have brought the cloud in such position. For example, the cloud might have achieved this orbit after a collision with another cloud. On the other hand, the existence of two giant molecular clouds, M$-0.02-0.07$ and M$-0.13-0.08$, at $\sim{}7$ and $\sim{}13$ pc, respectively, from the GC (Solomon et al. 1972; Okumura et al. 1991; Ho et al. 1991; Novak et al. 2000) shows that molecular clouds, probably unbound, exist at $\lesssim{}20$ pc from the GC. 
The cloud is assumed to be isothermal. This assumption neglects the local variations of the effective equation of state, that might occur due to the fact that the balance between heating and cooling processes depends on local conditions (see Section~4). 
In this Letter we focus on this case, as we want to present, in a concise way, our basic idea, preliminary simulations and findings. In a forthcoming paper we will carry out  a parametric study considering different initial positions and velocities, different masses and a varying internal structure 
of the cloud, as well as different equations of state.


SF in the cloud is allowed by means of sink particles, as in NCS07. We use an upgraded version of the parallel N-body/SPH code GASOLINE (Wadsley, Stadel \& Quinn 2004) in which sink particles have been implemented according to criteria widely used in the literature (Bate, Bonnell \& Price 1995). We adopt a sink accretion radius $r_{acc}=5\times{}10^{-3}$ pc, similar to the softening length ($\epsilon{}=5\times{}10^{-3}$ pc). 
We calculate a density threshold $\rho{}_{thr}= \Omega{}\,{}c_s /(2\,{}\pi{}\,{}G\,{}h\,{}Q)\simeq{} 10^6$ cm$^{-3}$ (where $\Omega{}$ is the angular velocity, $c_s$ the sound speed, $G$ the gravitational constant, $h$ the scaleheight of the disc and $Q$ is the Toomre parameter), by imposing that $Q=1$, according to Toomre's criterion for gravitational 
stability appropriate for rotating gaseous discs. Gas parcels that are unstable based on the Toomre criterion are always well resolved thanks to our mass resolution.
\footnote{The exact value of the density threshold is not crucial for our findings since values of $\rho{}_{thr}$ between $10^3$ and $10^8$ cm$^{-3}$ give approximately the same results. This is due to the fact that, given the high densities reached in the disc around the SMBH, the density threshold is easily reached, whereas the other criteria are more 
difficult to satisfy.} (Toomre 1964). Whenever a gas particle reaches this density threshold and its smoothing length is less than $0.5\,{}r_{acc}$ (so that at least $\sim{}$50 particles are inside $r_{acc}$ - for comparison in GASOLINE a single SPH kernel comprises 32 particles), it is considered a 'sink candidate'. If gas particles inside the accretion radius of the sink candidate satisfy Bate's criteria\footnote{Bate's criteria for converting gas particles into sink particles require (see section 2.2.2 of Bate et al. 1995) i) that the thermal energy of particles inside $r_{acc}$ is  $E_{th}\le{}0.5\,{}E_g$, where $E_g$ is the magnitude of the gravitational energy of the particles; ii) that $E_{th}/E_g + E_r/E_g\le{}1$, where $E_r$ is the rotational energy of particles; iii) that the total energy of particles is negative.},
 then the candidate becomes a sink particle. A similar procedure is followed to decide whether a particle which has already become a sink will accrete gas particles\footnote{According to Bate's criteria (see section 2.2.1 of Bate et al. 1995) a gas particle within $r_{acc}$ will be accreted by the sink particle if i) the gas particle is bound to the sink; ii) the specific angular momentum of the particle about the sink is less than required to form a circular orbit at $r_{acc}$; iii) the gas particle is more tightly bound to the considered sink particle than to other sink particles.}.
With this method we should be able to turn gas into stars according to the local properties of the cloud.
This approach has many intrinsic limits (e.g. we do not follow the fragmentation of the cloud, but we replace this process with the formation of sink particles), but it is one of the best available approximations, up to date (see also NCS07). Furthermore, in the next section we will show that the results derived from the sink method are consistent with a different analysis based on Toomre's criteria for stability.

\section{Results}
The cloud, orbiting around SgrA$^\ast{}$,  rapidly ($t\lesssim{}10^5$ yr) stretches towards the SMBH and is partially disrupted. The branch of the cloud which points toward the SMBH 
begins to spiral in 
towards the GC (see Fig.~\ref{fig:fig1}).
A dense, rotating gaseous disc forms at $t\sim{}3-3.5\times{}10^5$ yr at the location of the in-spiraling branch. Its initial density is $\sim{}1-5\times{}10^4$ cm$^{-3}$ (assuming molecular weight $\mu{}=2$), its outer radius is $r_{out}\sim{}0.5$ pc and its initial mass is $\sim{}330-1230\,{}M_\odot{}$. 
Such disc
is not an homogeneous disc but is the assembly of many concentric annuli, which spiral around the SMBH, with slightly different inclination and thickness (lower panel of Fig.~\ref{fig:fig1}). 
The initial average thickness of the disc, defined as the ratio between the average scaleheight $h$ and $r_{out}$, is $\sim{}0.1$. Many spiral perturbations can be also seen in the disc at this stage.

At $t\sim{}4\times{}10^5$ yr the disc has the average density of $\sim{}2\times{}10^5$ cm$^{-3}$, but 
local densities of $\sim{}10^{7-8}$ cm$^{-3}$ are 
reached (see the density map of gas in Fig.~\ref{fig:fig2}). At this stage, the total mass of the gaseous disc is $\sim{}2800-3100\,{}M_\odot{}$ and the parent cloud is still feeding it through a finger-like structure. 
The outer radius of the disc is still $r_{out}\sim{}0.5$ pc and it does not change during the entire simulation. 
The disc appears distorted at the edges, where fresh gas is being fed by the parent cloud.
Similarly, the orbits of gas particles are quite eccentric on the periphery of the disc ($e\lesssim{}0.5$) and almost circular at the centre ($e\lesssim{}0.1$).

Stars begin to form inside the disc at $t\gtrsim{}3.3\times{}10^5$ yr, i.e. immediately after the disc itself arises. Most of SF takes place between  $3.5$ and $5.0\times{}10^5$ yr. Between $3.3$ and $4.0\times{}10^5$ yr 55  stars form in the disc. In the next $10^5$ yr (i.e. between $4.0$ and $5.0\times{}10^5$ yr) other 101 new stars form, and the total number of stars in the disc reaches 156. The total mass in stars at $t=5.0\times{}10^5$ yr is $\sim{}4900\,{}M_\odot{}$. 
Figs.~\ref{fig:fig2} and ~\ref{fig:fig3} show the position of stars (black filled circles) superimposed to the density map of gas in the central disc, at $t=4\times{}10^5$ yr and $t=5\times{}10^5$ yr, respectively. As stars have formed inside the disc, the distribution of stars is also confined into a disc. The thickness of the stellar disc is $\sim{}0.05-0.08$ and does not evolve significantly during the simulation. The outer radius of the stellar disc is $r_{out}\sim{}0.5$ pc and does not change during the simulation. Stars which are corotating with the disc form only within $\sim{}0.5$ pc, as the parent gaseous disc does not extend beyond $\sim{}0.5$ pc. Only few stars have formed outside this radius (because the density rapidly drops below $\rho{}_{thr}$) and they are not corotating with the disc. 
This is in good agreement with the observations (G03; P06).
Interestingly, the stellar disc has also an inner radius ($r_{in}$): no stars have  formed inside $r_{in}\sim{}0.06$ pc, as it can be seen from the left-hand panels of Figs.~\ref{fig:fig2} and ~\ref{fig:fig3}. This  is in agreement with observations, which indicate that the stellar disc has a well-defined inner radius $r_{in}\sim{}0.04$ pc. 
In the simulation, the existence of the inner radius is probably due to the fact that even the high central density of gas cannot counteract the Keplerian velocity at such small distances from the SMBH. In fact, Toomre's $Q$ parameter, defined as $Q=\Omega{}\,{}c_s/(\pi{}\,{}G\,{}\Sigma{})$ (where 
$\Sigma{}$ is the local surface density, Toomre 1964), is $Q\gtrsim{}5$ at radii $\lesssim{}0.04$ pc. For such high values of $Q$, the growth of gravitational instabilities in the disc is unlikely. Thus, a very dense ($\sim{}10^8$ cm$^{-3}$) gaseous disc, with a radius $\sim{}0.04$ pc, survives around the SMBH during the entire simulation.
On the other hand, although the softening length is $\sim{}10$ times smaller than $r_{in}$, we cannot completely exclude that the existence of $r_{in}$ is due to numerical effects.
Higher resolution simulations are needed to address this issue.
The comparison between Fig.~\ref{fig:fig2} and Fig.~\ref{fig:fig3} reveals another interesting feature: gas is being gradually depleted from the inner regions of the star forming disc. In fact, at $t=5\times{}10^5$ yr (Fig.~\ref{fig:fig3})  the density of gas between $\sim{}0.05$ and $\sim{}0.25$ pc is much lower than at $t=4\times{}10^5$ yr (Fig.~\ref{fig:fig2}).
Correspondingly, the number of stars between $\sim{}0.05$ and $0.25$ pc in Fig.~\ref{fig:fig3} is a factor of $\sim{}2$ higher than in Fig.~\ref{fig:fig2}. Thus, gas is depleted from the inner parts of the disc, because it has been efficiently converted into stars. Second, the feeding from the parent cloud tends to decrease, and is not able to counter-balance gas consumption by SF.

After $5.0\times{}10^5$ yr, the SF rapidly declines. In fact, most of the densest gas within $r_{out}\sim{}0.5$ pc has been converted into stars, and the remaining gas is not sufficiently dense to produce new stars. 
At $t=6.0\times{}10^5$ yr the total stellar mass  within the disc is $\sim{}5850\,{}M_\odot{}$, and no significant changes have occurred either in the stellar distribution or in the mass function (MF) during the last $10^5$ yr. At the end of the simulation ($t=7.0\times{}10^5$ yr) the total stellar mass within the disc is still $\lesssim{}6000\,{}M_\odot{}$, indicating that the SF process has almost stopped. This total mass is quite in agreement with the estimated value of $\sim{}5000\,{}M_\odot{}$ and is well below the upper limit of $\sim{}1.5\times{}10^4\,{}M_\odot{}$ derived from observations of the discs around SgrA$^{\ast{}}$ (P06). As the parent gaseous disc was eccentric ($e\lesssim{}0.5$) and was distorted in its outer parts, the stellar orbits  have non-negligible eccentricities with a large spread ($0.1\lesssim{}e\lesssim{}0.7$) and different inclinations with respect to the disc plane (between  $\sim{}0$ and  $\sim{}0.7$ rad). This result is consistent with recent observations (P06; Cuadra, Armitage, Alexander 2008).
 At this stage, the entire parent cloud is stretched and very elongated toward SgrA$^\ast{}$, its average density is $\sim{}55$ cm$^{-3}$ and its centre-of-mass is located at 
 $\sim{}10$ pc from the GC. These characteristics, and especially the elongation toward SgrA$^\ast{}$, are quite similar to those of the molecular cloud M$-0.13-0.08$. The average accretion rate of the SMBH during the simulation is $\sim{}5\times{}10^{-6}M_\odot{}\textrm{ yr}^{-1}$, below the  gas capture rate inferred from X-ray observations ($10^{-5} M_\odot{}\textrm{ yr}^{-1}$; Baganoff et al. 2003) but a factor of 5 higher than the upper limit from polarization measurements (Marrone et al. 2006). However, this value must be considered a rough approximation, as we cannot describe the physics of the accretion around the SMBH.




\begin{figure}
\center{{
\epsfig{figure=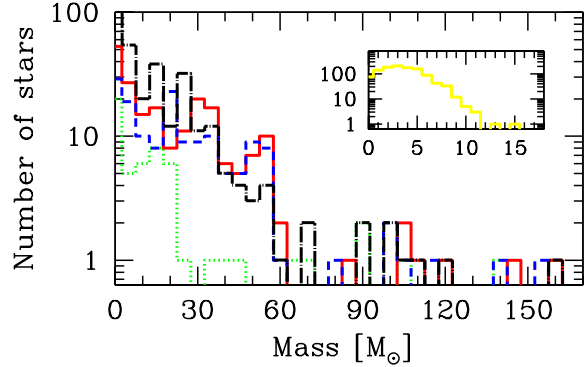,width=8.5cm}
}}
\caption{\label{fig:fig4} 
Stellar MF in the simulation. Dotted green line: $t=4\times{}10^5$ yr; dashed blue line: $t=5\times{}10^5$ yr; solid red line: $t=6\times{}10^5$ yr. Dot-dashed black line: stellar MF in the simulation at $t=6\times{}10^5$ yr when accounting for binary fraction. 
In the small insert: solid yellow line: MF in the isolated cloud ($t=16$ Myr).
}
\end{figure}

Fig.~\ref{fig:fig4} shows the stellar MF derived from our simulation at $t=4$, 5 and $6\times{}10^5$ yr (dotted, dashed and solid line, respectively). After a fast initial evolution ($t<5\times{}10^5$ yr), the MF does not change significantly between 5 and $6\times{}10^5$ yr. 
In the final MF most of stars have mass below $60\,{}M_\odot{}$. 
The number of OB and Wolf-Rayet stars detected in the GC up to date is $\sim{}73$,  and their inferred total mass is $\sim{}3700-4300\,{}M_\odot{}$, assuming a mass range of $20-120\,{}M_\odot{}$ and an IMF $dN/dm\propto{}m^\Gamma{}$ with slope $\Gamma{}$ between $-1.35$ and $-0.85$ (P06).
In our simulation 
there are 90 stars in the same range of mass ($20-120\,{}M_\odot{}$), corresponding to a total mass of $\sim{}3740\,{}M_\odot{}$.
The agreement between data and simulation is quite good.
A possible problem of the simulated IMF is the existence of a very high-mass tail. 13 stars in the simulations have mass higher than 60 $M_\odot{}$, and the most massive among them has $m=202\,{}M_\odot{}$. The formation of excessively massive stars may be an intrinsic problem of the sink particle method. In fact, the use of sink particles does not allow to resolve close binary systems (Klessen, Spaans \& Jappsen 2007), but the fraction of binaries is known to be $\sim{}0.5$, at least in the solar neighborhoods (Vanbeveren, De Loore \& Van Rensbergen 1998). To account for this, we assume that half of the stars in the simulation are unresolved binaries, with equal mass components, and we Monte-Carlo sample the MF based on such assumption. 
The MF obtained 
with this procedure is shown by the dot-dashed black line in  Fig.~\ref{fig:fig4}. In this case, the most massive star weighs  $m=161\,{}M_\odot{}$.

Nevertheless, the properties of the gaseous disc are consistent with the formation of massive stars. In order to show that, we have calculated the Toomre most unstable wavelength 
($\lambda{}_{\rm mu}=2\,{}\pi{}^2\,{}G\,{}\Sigma{}/\kappa{}^2$, 
 where $\kappa{}$ is the epicyclic frequency, Binney \& Tremaine 1987), i.e. the wavelength at which instability first appears, when $Q$ drops below unity in a differentially rotating disc.  We find that, at $t=3.9\times{}10^5$ Myr, $\lambda{}_{\rm mu}$ can be as large as $\sim{}4.1\times{}10^{16}$ cm (at $\sim{}0.37$ pc). The mass enclosed into a spherical volume of radius $\lambda{}_{\rm mu}$, $\sim{}30\,{}M_\odot{}$, represents then the expected characteristic mass at which collapse takes place and nicely agrees with the typical stellar mass found in the simulations.
For a further check, we also ran a simulation in which the cloud is isolated (i.e. SMBH and rigid potential are not present) and we derived the corresponding MF. In this case, SF starts much later ($t\sim{}3$ Myr) and reaches a maximum at $t\sim{}12$ Myr, approximately the dynamical time of the cloud. No stars with mass higher than $\sim{}15\,{}M_\odot{}$ form in the isolated cloud (see insert in Fig.~\ref{fig:fig4}). Thus, we conclude that it is the dynamical interaction between the SMBH and the cloud, with the resulting formation of the disc,
that triggers a top-heavy IMF.

\section{Conclusions}
We simulated the infall of a molecular cloud toward SgrA$^\ast{}$. In the first $\sim{}10^5$ yr the cloud is disrupted by the tidal forces of the SMBH and starts spiraling towards it. At $t\sim{}3\times{}10^5$ yr a dense and small ($\lesssim{}0.5$ pc) gaseous disc forms around the SMBH. Due to the high densities reached by the gas, stars begin to form in the disc. At $t\sim{}5-6\times{}10^5$ yr most of gas has been depleted from the disc and converted into stars. The stars are distributed in a thin disc with $r_{in}\sim{}0.06$ and $r_{out}\sim{}0.5$ pc, similar to the one observed around SgrA$^\ast{}$ (G03; P06). The total stellar mass in the disc, $\sim{}6000\,{}M_\odot{}$, is also in agreement with observations. We found that the MF of the simulated stellar disc is top-heavy. 
A simple  estimate based on Toomre's most unstable wavelength predicts the formation of massive stars in the gaseous disc, in agreement with the results obtained from the sink particle method. Thus, our simulations suggest that the infall of a molecular cloud toward the GC is a viable scenario for the formation of the massive young stars around SgrA$^\ast{}$. This is an important result, as the origin of the stellar disc around SgrA$^\ast{}$ was still a puzzle so far.

However, our method suffers from various limitations and assumptions. For example, the timescale of SF likely depends on the assumption of isothermal equation of state. We expect longer timescales for SF when, e.g., a  polytropic equation of state with a variable adiabatic index $\gamma{}$ is adopted, because the gas would become more pressurized against collapse.
However, even a factor of $\sim{}5$ longer timescale is in agreement with observations, which suggest that the formation of the stellar disc took $<2$ Myr. Even the stellar MF might 
depend on the treatment of gas thermodynamics and on the recipe to initialize sink particles. 
In a forthcoming paper (Mapelli et al., in preparation) we will study the dependence of our results on the equation of state of the gas.
Furthermore, only one stellar disc forms during this simulations, whereas observations suggest the existence of two different discs. It may be possible that two different clouds have produced two different discs with similar ages, or that the same inspiraling cloud has fed two discs at slightly different epochs.
In conclusion, this paper represents the first attempt to understand the formation of massive stars around SgrA$^\ast{}$ by simulating directly the infall of a molecular cloud. 
The results are encouraging, but this scenario deserves further investigation with a  more realistic model of the thermodynamics and SF in the interstellar cloud and in the disc.


\section*{Acknowledgments}
We thank B. Moore, E.~D'Onghia, E. Ripamonti, S. Callegari and P.~Englmaier for useful discussions.
MM, TH and LM   acknowledge support from the Swiss
National Science Foundation.


{}


\begin{thebibliography}{}

\bibitem{}Alexander R. D., Armitage P. J., Cuadra J., Begelman M. C., 2008, ApJ, 674, 927

\bibitem{}Baganoff F. K. et al., 2003, ApJ, 591, 891

\bibitem{}Bate M. R., Bonnell I. A., Price N. M., 1995, MNRAS, 277, 362

\bibitem{}Binney J., Tremaine S., 1987, {\it Galactic dynamics}, Princeton University Press

\bibitem{}Collin S., Zahn J.-P.,  2008, A\&{}A, 477, 419

\bibitem{}Cuadra J., Armitage P. J., Alexander R. D., 2008, MNRAS, submitted



\bibitem{}Eisenhauer F., et al., 2005, ApJ, 628, 246

\bibitem{}Fujii M., Iwasawa M., Funato Y., Makino J., 2008, submitted to ApJ, arXiv:0708.3719

\bibitem{}Genzel R., et al. 2003, ApJ, 594, 812 (G03)

\bibitem{}Gerhard O., 2001, ApJ, 546, L39

\bibitem{}Ghez A. M. et al., 2003ApJ, 586L, 127

\bibitem{}Ghez A. M., Salim S., Hornstein S. D., Tanner A., Morris M., Becklin E. E., Duchene G., 2005, ApJ, 620, 744

\bibitem{}Goodman J., 2003, MNRAS, 339, 937

\bibitem{}G\"urkan M. A.,  Rasio F. A., 2005, ApJ, 628, 236


\bibitem{}Ho P. T. P., Ho L. C., Szczepanski J. C., Jackson J. M., Armstrong J. T., Barrett A. H., 1991, Nature, 350, 309



\bibitem{}Kim S. S.,  Morris M, 2003, ApJ, 597, 312

\bibitem{}Klessen R. S., Spaans M., Jappsen A.-K., 2007, MNRAS, 374L,~29

\bibitem{}Larson R. B., 1981, MNRAS, 194, 809

\bibitem{}Levin Y.,  Beloborodov A. M., 2003, ApJ, 590L, 33

\bibitem{}Marrone D. P., Moran J. M., Zhao J.-H., Rao R., 2006, ApJ, 640, 308

\bibitem{}McMillan S. L. W.,  Portegies Zwart S. F., 2003, ApJ, 596, 314

\bibitem{}Milosavljevic M., Loeb A., 2004, ApJ, 604L, 45


\bibitem{}Nayakshin S., Cuadra J. J., 2005, A\&{}A, 437, 437

\bibitem{}Nayakshin S., Cuadra J., Springel V., 2007, MNRAS, 379, 21 (NCS07)

\bibitem{}Nayakshin S., Sunyaev R., 2005, MNRAS, 364L, 23

\bibitem{}Novak G., Dotson J. L., Dowell C. D., Hildebrand R. H., Renbarger T., Schleuning D. A., 2000, ApJ, 529, 241

\bibitem{}Okumura S. K., Ishiguro M., Fomalont E. B., Hasegawa T., Kasuga T., Morita K. I., Kawabe R.,  Kobayashi H., 1991, ApJ, 378, 127

\bibitem{}Paumard T. et al., 2006, ApJ, 643, 1011 (P06)

\bibitem{}Portegies Zwart S. F., Baumgardt H., McMillan S. L. W., Makino J., Hut P., Ebisuzaki T., 2006, ApJ, 641, 319


\bibitem{}Solomon P. M., Scoville N. Z., Jefferts K. B., Penzias A. A., Wilson R. W., 1972, ApJ, 178, 125

\bibitem{}Spaans M., Silk J., 2000, ApJ, 538, 115


\bibitem{}Toomre A., 1964, ApJ, 139, 1217

\bibitem{}Vanbeveren D., De Loore C., Van Rensbergen W., 1998, A\&{}A Review, 9, 63

\bibitem{}Wadsley J. W., Stadel J., Quinn T., 2004, New Astronomy, 9, 137


\end{thebibliography}
\end{document}